\documentclass[12pt]{iopart}
\usepackage{cite} 
\usepackage{wrapfig}
\usepackage{graphicx}
\usepackage{graphics}
\usepackage{hyperref}

\begin{document}

\title[The exact solution of a  generalized Bose-Hubbard model]{The exact solution of a generalized Bose-Hubbard model}

\author{ Gilberto N. Santos Filho$^{1}$}

\address{$^{1}$ Centro Brasileiro de Pesquisas F\'{\i}sicas - CBPF \\ 
Rua Dr. Xavier Sigaud, 150, Urca, Rio de Janeiro - RJ - Brazil}
\ead{gfilho@cbpf.br}


\begin{abstract}
I present the exact solution of a family of fragmented Bose-Hubbard models and represent the models as graphs in one-dimension, two-dimensions and three-dimensions with the condensates in the vertices. The models are solved by the algebraic Bethe ansatz method.
\end{abstract}

\section{Introduction}

  The realization of Bose-Einstein condensates (BEC)\cite{bose,einstein}, achieved by taking dilute alkali gases to ultra low temperatures \cite{early,angly,wwcch,anderson,wk,Hulet} 
is certainly among the most exciting recent experimental achievements in physics. Since then, investigations dedicated to the comprehension 
of  new phenomena associated to this state of matter as well as its properties have flourished, either in the experimental or theoretical domains. 
 The fast increasing of the control and production of Bose-Einstein condensates (BECs) in different geometries has permitted the study of these systems in different physical situations. The fragmentation of a BEC to produce a Josephson junction \cite{Josephson1,Josephson2} using BECs opened the possibility to study the quantum tunnelling of the atoms across a barrier between the condensates \cite{Ketterle1,Ketterle2,Oberthaler1,Oberthaler2,Oberthaler3}.   Using superposition of light in different direction is possible to create any arbitrary trapping configuration as for example a ring or a superconducting quantum interference devices (SQUID) with an  atom BEC \cite{Ryu1,Ryu2}. Another experimental realization of fragmentation of BECs is the two-legs bosonic ladder to study chiral current and Meissner effect \cite{ladder1,ladder2}. There is also the possibility to produce optical lattices in one-dimension (1D), two-dimensions (2D) and three-dimensions (3D) using one, two or three orthogonal standing waves \cite{Bloch}. These experimental realisations have opened the possibility to introduce new models that permit the study of the tunnelling of the atoms between the BECs. Some of these models are exactly solvable by the algebraic Bethe ansatz method \cite{Roditi,jon1,jon2,jonjpa,key-3,dukelskyy,Ortiz,Kundu,eric5,GSantosaa,GSantos,GSantos11,multistates,Xin,Ymai,Bethe-states,rubeni} and this open the possibility to take in account quantum fluctuations that allows us go beyond the results obtained by mean field approximations. This fruitful approach can furnish some new insights in 
this area, and contribute as well to the increasingly interesting field of integrable systems itself \cite{hertier, batchelor1,batchelor2,batchelor3}. The algebraic formulation of the Bethe ansatz, and the associated quantum inverse scattering method (QISM), was primarily  developed in \cite{fst,ks,takhtajan,korepin,faddeev}. The QISM has been used to unveil properties of a considerable number of  solvable systems, such as, one-dimensional spin chains, quantum field theory of one-dimensional 
interacting bosons \cite{korepin1} and fermions \cite{yang}, two-dimensional lattice models \cite{korepin2}, systems of strongly correlated electrons \cite{ek,ek2}, conformal field theory \cite{blz}, integrable systems in high energy physics \cite{lipatov, korch, belitsky,Gromov1,Gromov2,CAhn}
 and quantum algebras (deformations of universal
enveloping algebras of Lie algebras) \cite{jimbo85,jimbo86,drinfeld,frt}.  For a pedagogical and historical review see \cite{faddeev2}. More recently solvable 
models have also showed up in relation to string theories (see for instance \cite{dorey}).  Remarkably it is important to mention that exactly solvable models 
are recently finding their way into the lab, mainly in the context of ultracold atoms \cite{batchelor2} but also in nuclear magnetic resonance (NMR) experiments\cite{screp,kino1,kino2,kitagawa,haller,liao,coldea,nmr1}
turning its study as well as the derivation of 
new models an even more fascinating field.

I am considering here the bosonic multi-states Lax operator introduced by the author  in \cite{multistates} that permits to solve a family of models of fragmented BECs coupled by Josephson tunnelling. This Lax operator is a generalization of the bosonic Lax operator in \cite{Kuznet, key-3}, where a Lax operator is 
defined for a single canonical boson operator, but instead of a single operator we choose a linear combination of independent canonical boson operators.

The paper is organized as follows. In section 2, I will review briefly the algebraic
Bethe ansatz method and present the Lax operators and the transfer matrix for both
models. In section 3, I present a generalized model with a  family of models of fragmented BECs coupled by Josephson tunnelling and show their solutions. In section 4, I summarize the results.

\section{Algebraic Bethe ansatz method}

In this section we will shortly review the algebraic Bethe ansatz method and present the transfer matrix used to get the solution of the 
models \cite{jonjpa,Roditi}. We begin with the $gl(2)$-invariant $R$-matrix, depending on the spectral parameter $u$,

\begin{equation}
R(u)= \left( \begin{array}{cccc}
1 & 0 & 0 & 0\\
0 & b(u) & c(u) & 0\\
0 & c(u) & b(u) & 0\\
0 & 0 & 0 & 1\end{array}\right),\end{equation}

\noindent with $b(u)=u/(u+\eta)$, $c(u)=\eta/(u+\eta)$ and $b(u) + c(u) = 1$. Above,
$\eta$ is an arbitrary parameter, to be chosen later.

It is easy to check that $R(u)$ satisfies the Yang-Baxter equation

\begin{equation}
R_{12}(u-v)R_{13}(u)R_{23}(v)=R_{23}(v)R_{13}(u)R_{12}(u-v),
\end{equation}

\noindent where $R_{jk}(u)$ denotes the matrix acting non-trivially
on the $j$-th and the $k$-th spaces and as the identity on the remaining
space.

Next we define the monodromy matrix  $\hat{T}(u)$,

\begin{equation}
\hat{T}(u)= \left( \begin{array}{cc}
 \hat{A}(u) & \hat{B}(u)\\
 \hat{C}(u) & \hat{D}(u)\end{array}\right),\label{monod}
\end{equation}

\noindent such that the Yang-Baxter algebra is satisfied

\begin{equation}
R_{12}(u-v)\hat{T}_{1}(u)\hat{T}_{2}(v) = \hat{T}_{2}(v)\hat{T}_{1}(u)R_{12}(u-v).\label{RTT}
\end{equation}

\noindent In what follows we will choose a realization for the monodromy matrix $\pi(\hat{T}(u))=\hat{L}(u)$  
to obtain solutions of a family of models for multilevel two-well Bose-Einstein condensates.
In this construction, the Lax operators $\hat{L}(u)$  have to satisfy the relation

\begin{equation}
R_{12}(u-v)\hat{L}_{1}(u)\hat{L}_{2}(v)=\hat{L}_{2}(v)\hat{L}_{1}(u)R_{12}(u-v).
\label{RLL}
\end{equation}

Then, defining the transfer matrix, as usual, through

\begin{equation}
\hat{t}(u)= tr \;\pi(\hat{T}(u)) = \pi(\hat{A}(u) + \hat{D}(u)),
\label{trTu}
\end{equation}
\noindent it follows from (\ref{RTT}) that the transfer matrix commutes for
different values of the spectral parameter; i. e.,

\begin{equation}
[\hat{t}(u),\hat{t}(v)]=0, \;\;\;\;\;\;\; \forall \;u,\;v.
\end{equation}
\noindent Consequently, the models derived from this transfer matrix will be integrable. Another consequence is that the 
coefficients $\hat{\mathcal{C}}_k$ in the transfer matrix $\hat{t}(u)$,

\begin{equation}
\hat{t}(u) = \sum_{k} \hat{\mathcal{C}}_k u^k,
\end{equation}
\noindent are conserved quantities or simply $c$-numbers, with

\begin{equation}
[\hat{\mathcal{C}}_j,\hat{\mathcal{C}}_k] = 0, \;\;\;\;\;\;\; \forall \;j,\;k.
\end{equation}

If the transfer matrix $\hat{t}(u)$ is a polynomial function in $u$, with $k \geq 0$, it is easy to see that,

\begin{equation}
\hat{\mathcal{C}}_0 = \hat{t}(0) \;\;\; \mbox{and} \;\;\; \hat{\mathcal{C}}_k = \frac{1}{k!}\left.\frac{d^k\hat{t}(u)}{du^k}\right|_{u=0}. 
\label{C14b}
\end{equation}

 For the standard bosonic operators satisfying the canonical commutation relations 

\begin{equation}
 [\hat{p}_{i}^{\dagger},\hat{q}_{j}^{\dagger}] = [\hat{p}_{i},\hat{q}_{j}] = 0, \qquad [\hat{p}_{i},\hat{q}_{j}^{\dagger}] = \delta_{pq}\delta_{ij}\hat{I},
\end{equation}
\begin{equation}
[\hat{N}_{pi},\hat{q}_{j}^{\dagger}]= +\hat{p}_{j}^{\dagger}\delta_{pq}\delta_{ij}, \qquad [\hat{N}_{pi},\hat{q}_{j}]= -\hat{p}_{j}\delta_{pq}\delta_{ij},
\end{equation}
\noindent with $p,q = a \;\mbox{or}\; b$, $i = 1,\ldots, n$ and $j = 1, \ldots, m$, we have the following Lax operators,

\begin{equation}
\hat{L}^{\Sigma_{a}^n}(u) = \left(\begin{array}{cc}
u\hat{I} + \eta\sum_{j=1}^{n}\hat{N}_{aj} & \sum_{j=1}^{n}t_{aj}\hat{a}_{j}\\
\sum_{j=1}^{n}s_{aj}\hat{a}_{j}^{\dagger} & \eta^{-1}\zeta_a \hat{I}
\end{array}\right),
\label{L2}
\end{equation}
\noindent and
\begin{equation}
\hat{L}^{\Sigma_{b}^m}(u) = \left(\begin{array}{cc}
u\hat{I} + \eta\sum_{k=1}^{m}\hat{N}_{bk} & \sum_{k=1}^{m}t_{bk}\hat{b}_{k}\\
\sum_{k=1}^{m}s_{bk}\hat{b}_{k}^{\dagger} & \eta^{-1}\zeta_b \hat{I}
\end{array}\right),
\label{L3}
\end{equation}

\noindent if the conditions, $\zeta_a = \sum_{j=1}^{n}s_{aj}t_{aj}$ and $\zeta_b = \sum_{k=1}^{n}s_{bj}t_{bj}$, are satisfied. The above Lax operators satisfy the equation (\ref{RLL}).

\section{The Generalized Model}

In this section I present some applications of the Lax operators (\ref{L2}) and (\ref{L3}) for models with different number of BECs $a$ and $b$. The generalized Hamiltonian is,

\begin{eqnarray}
\hat{H} & = & \sum_{j=1}^{n} U_{ajaj}\hat{N}^2_{aj} + \sum_{k=1}^{m} U_{bkbk}\hat{N}^2_{bk}  + \frac{1}{2}\sum_{ j=1 }^{n}\sum_{ k=1 (j\neq k)}^{n} U_{ajak} \hat{N}_{aj}\hat{N}_{ak} \nonumber \\ 
&+& \frac{1}{2}\sum_{ j=1 }^{m}\sum_{ k=1 (j\neq k)}^{m} U_{bjbk} \hat{N}_{bj}\hat{N}_{bk}\nonumber \\ 
&+& \sum_{j=1}^{n}\sum_{k=1}^{m} U_{ajbk} \hat{N}_{aj}\hat{N}_{bk} - \sum_{j=1}^{n}\sum_{k=1}^{m} \mu_{ajbk}(\hat{N}_{aj} - \hat{N}_{bk}) \nonumber \\ 
&+& \sum_{j=1}^{n} \epsilon_{aj}\hat{N}_{aj} + \sum_{k=1}^{m} \epsilon_{bk} \hat{N}_{bk} -  \sum_{j=1}^{n}\sum_{k=1}^{m} J_{ajbk}(\hat{a}_{j}^{\dagger}\hat{b}_{k} + \hat{b}_{k}^{\dagger}\hat{a}_{j}).
\label{H1}
\end{eqnarray}
\noindent The parameters, $U_{pjqk}$, describe the atom-atom $S$-wave scattering between the atoms in the respective BECs, the $\mu_{ajbk}$ parameters are the relative external potentials between the BECs and  $\epsilon_{pj}$ are the energies in the BECs. The parameters $J_{ajbk}$ are the tunnelling amplitudes. The operators $N_{pj}$ are the number of atoms operators. The labels $p$ and $q$ stand for the BECs $a$ and $b$ with $j = 1,\ldots,n$ and $k = 1,\ldots,m$. We just remark that $U_{pjpk} = U_{pkpj}$. The BECs are coupled by Josephson tunnelling and the total number of atoms, $\hat{N} = \sum_{j=1}^n \hat{N}_{aj}  + \sum_{k=1}^m \hat{N}_{bk}$, is a conserved quantity, $[\hat{H},\hat{N}] = 0$.

\begin{figure}
\begin{center}
\begin{tabular}{cc}
$(a)$ & $(b)$ \\
\includegraphics[scale=0.5]{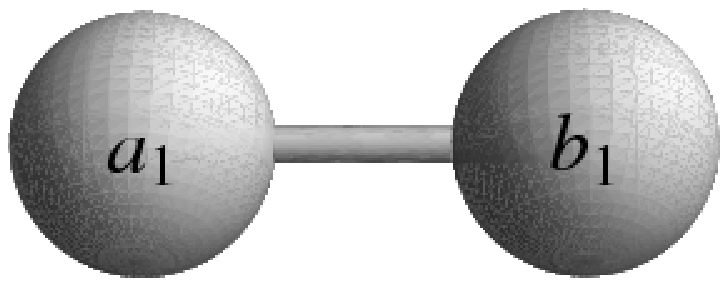} & \includegraphics[scale=0.5]{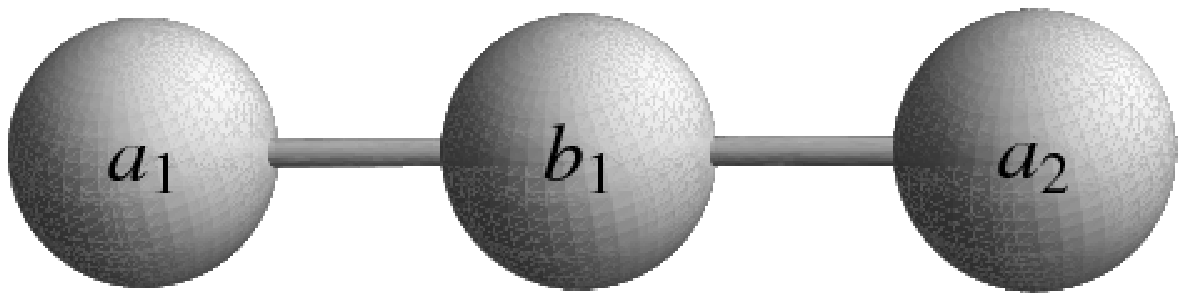} \\
$(c)$ & $(d)$ \\
\includegraphics[scale=0.5]{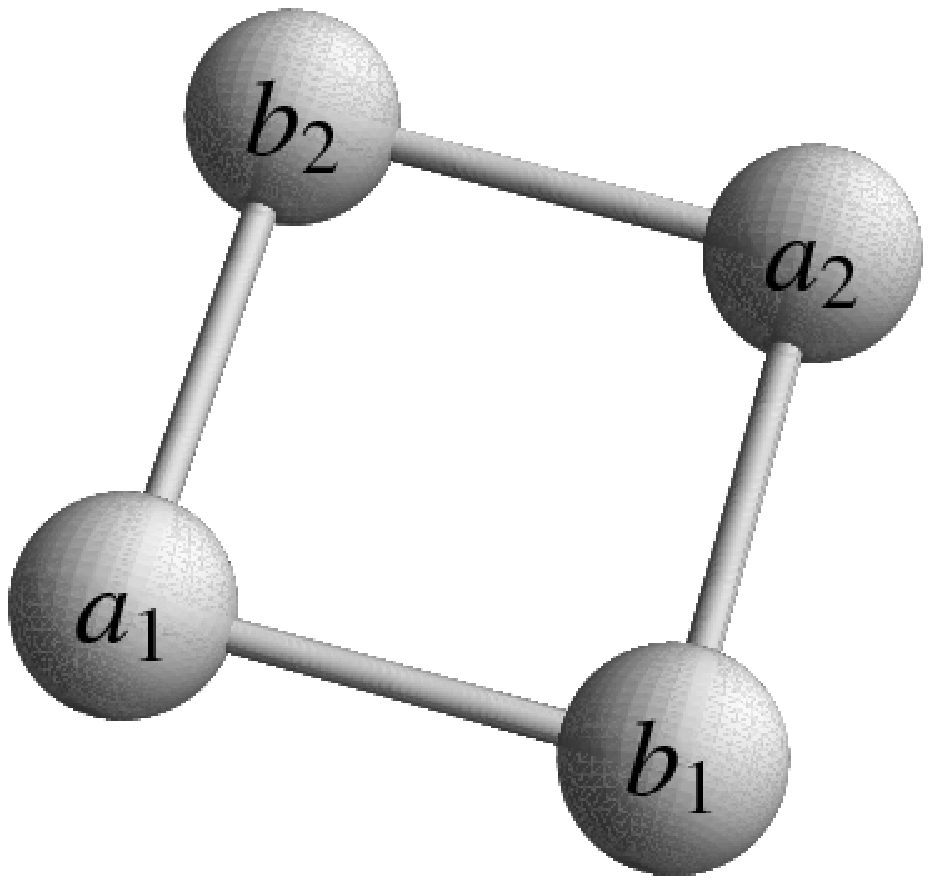} & \includegraphics[scale=0.5]{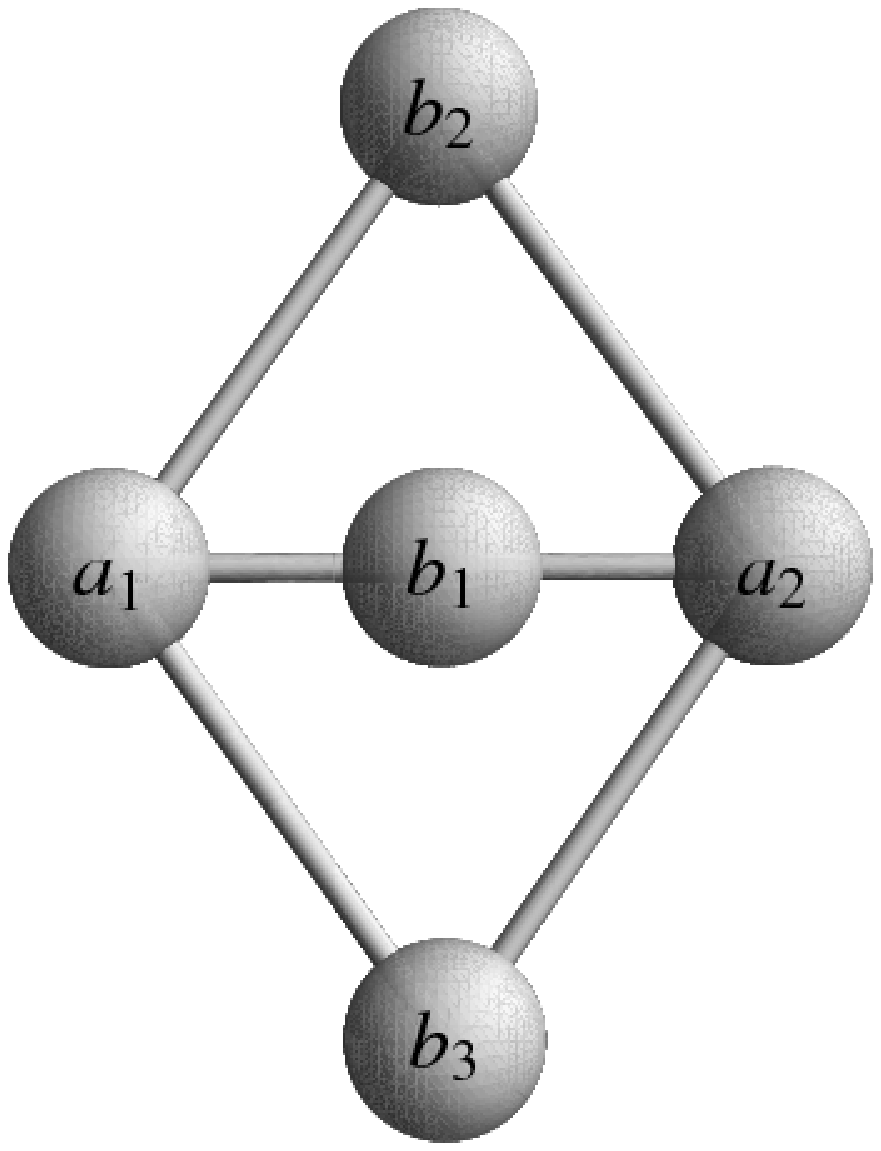} \\
$(e)$ & $(f)$ \\
\includegraphics[scale=0.5]{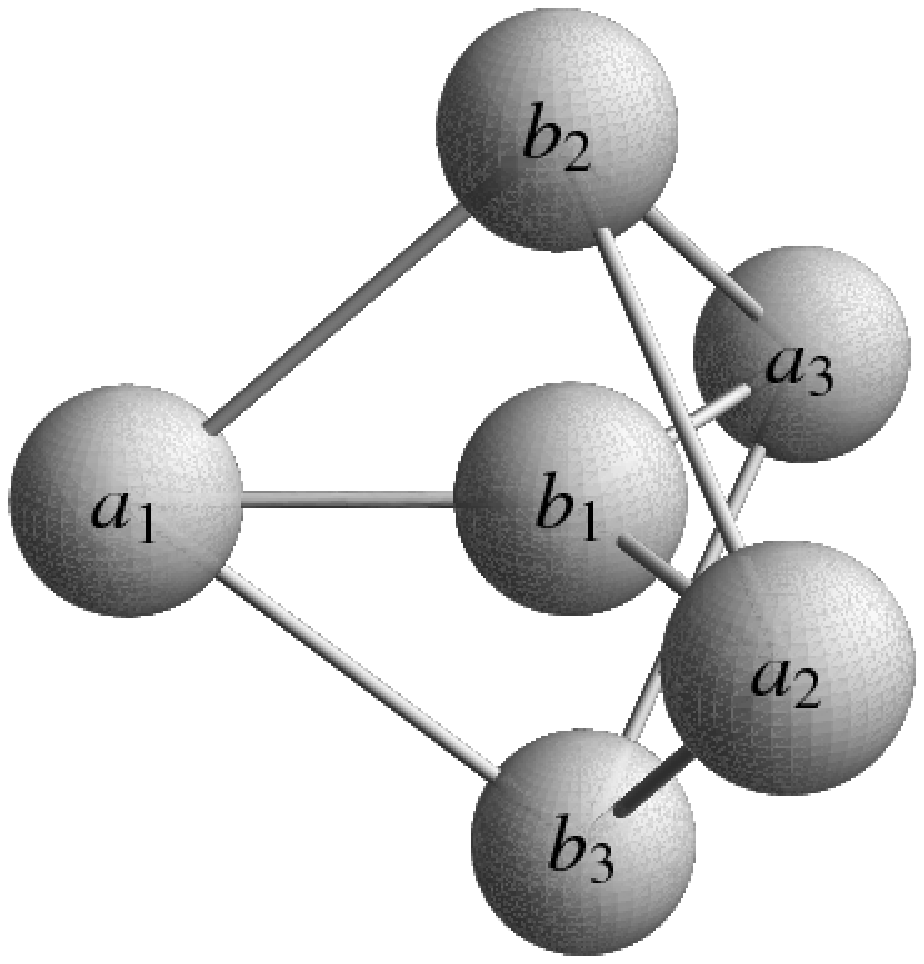} & \includegraphics[scale=0.5]{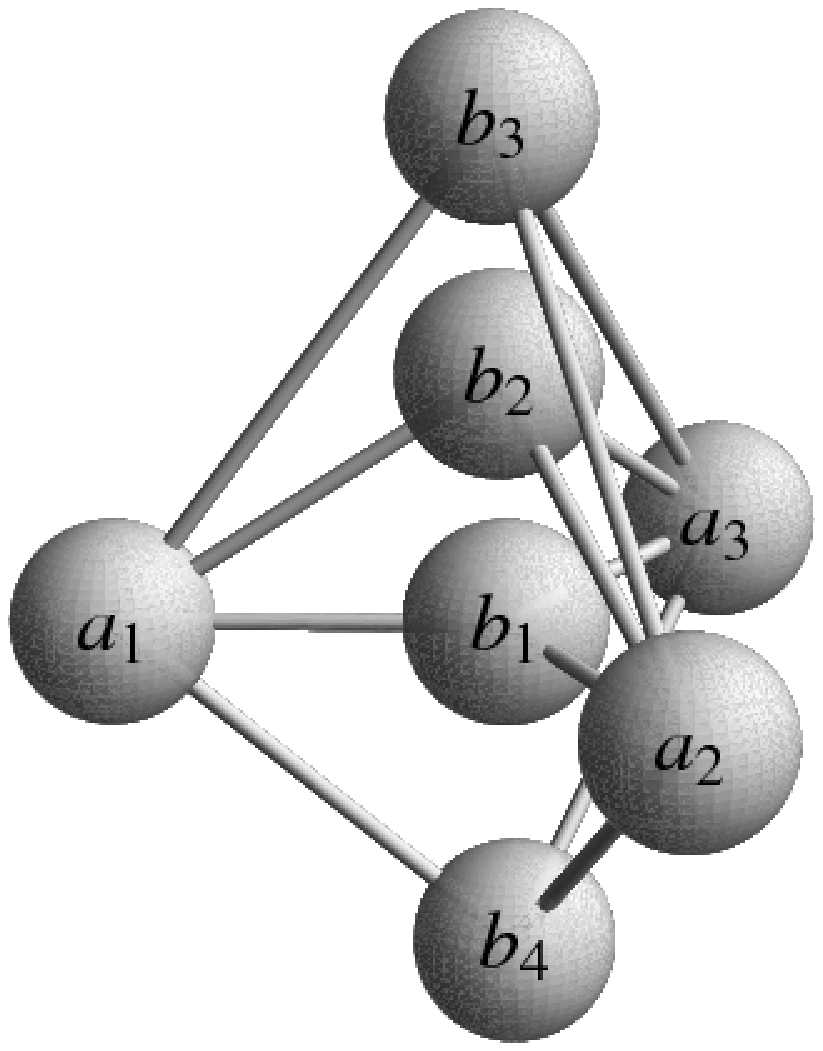} \\
\end{tabular}
\caption{In  $(a)$ we have $n = 1$ and $m=1$, in $(b)$ we have $n = 2$ and $m=1$, in $(c)$ $n = 2$ and $m=2$, in $(d)$ we have $n = 2$ and $m=3$, in $(e)$ we have $n = 3$ and $m=3$, in $(f)$ $n = 3$ and $m=4$.} 
\label{GF1}
\end{center}
\end{figure}

\begin{figure}
\begin{center}
\begin{tabular}{cc}
$(g)$ & $(h)$ \\
\includegraphics[scale=0.7]{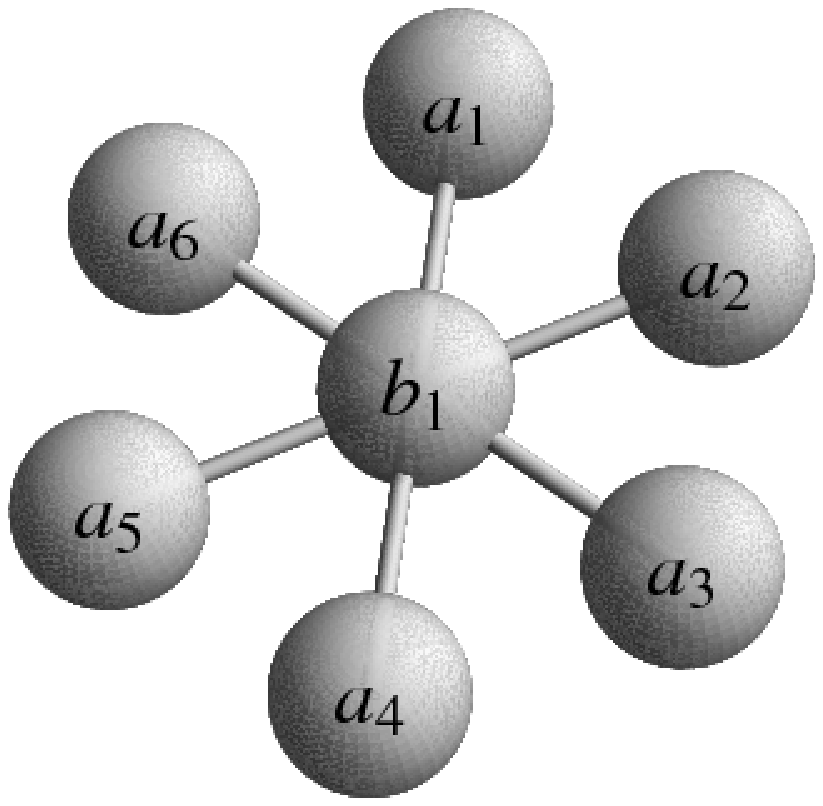} & \includegraphics[scale=0.5]{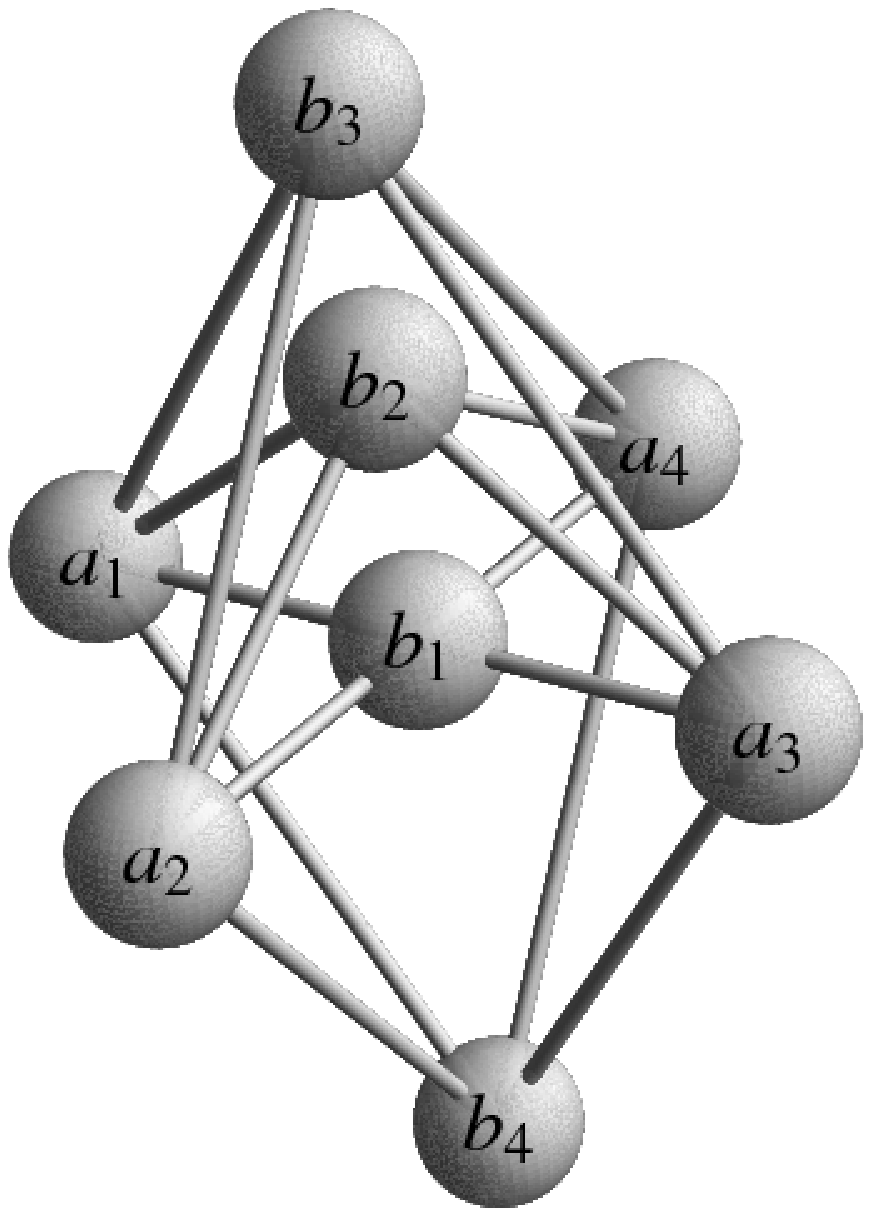} \\
$(i)$ & $(j)$ \\
\includegraphics[scale=0.5]{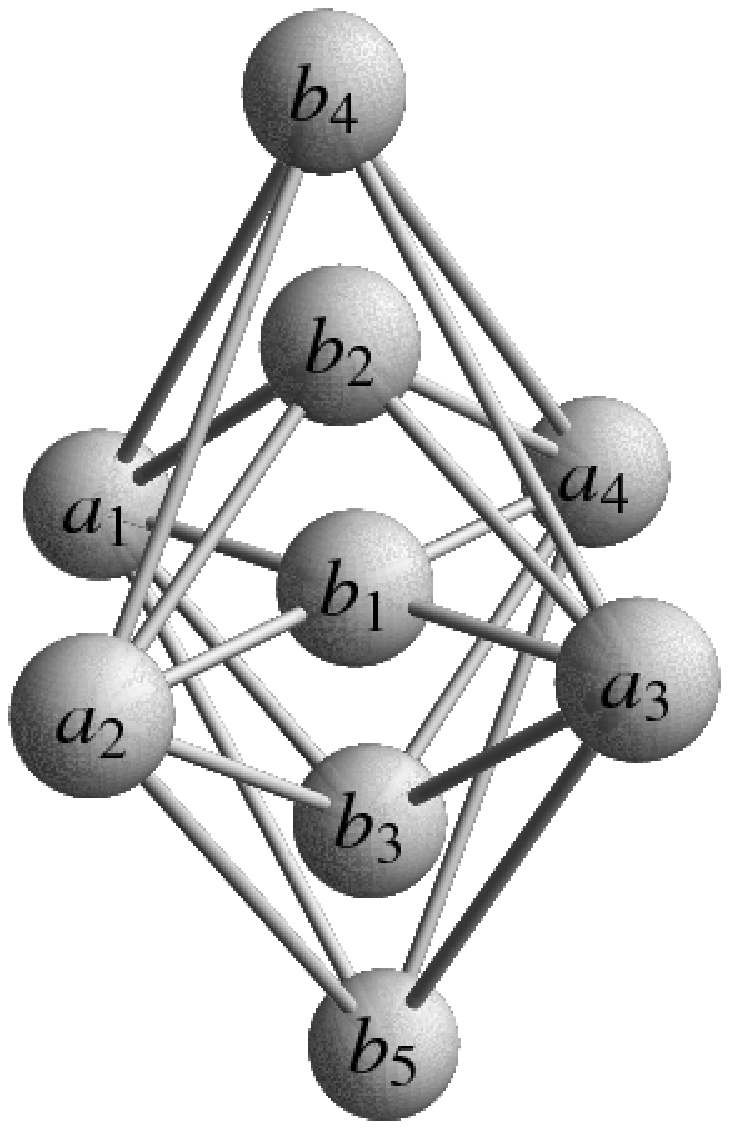} & \includegraphics[scale=0.5]{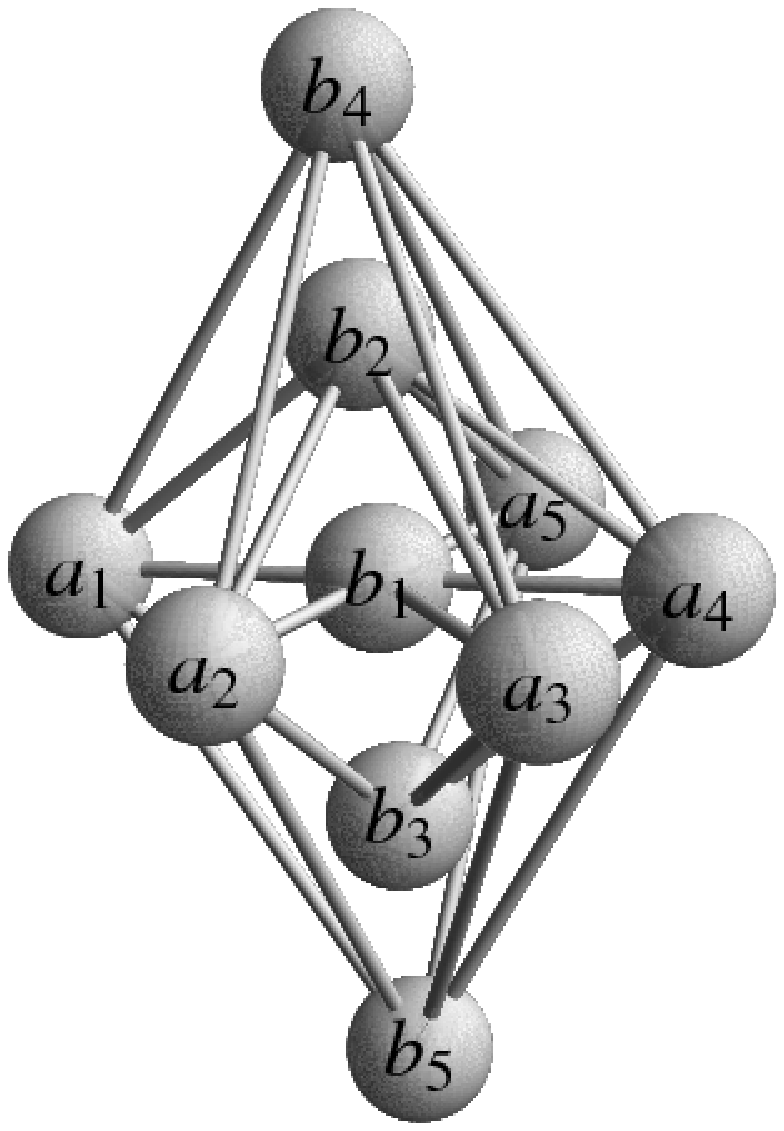} \\
\end{tabular}
\caption{In  $(g)$ we have $n = 6$ and $m=1$, in $(h)$ we have $n = 4$ and $m=4$, in $(i)$ $n = 4$ and $m=5$, in $(j)$ we have $n = 5$ and $m=5$.} 
\label{GF2}
\end{center}
\end{figure}

 The state space is spanned by the base $\{|n_{a1},\ldots,n_{an},n_{b1},\ldots,n_{bm}\rangle\}$ and we can write each vector state as 

\begin{eqnarray}
  |n_{a1},\ldots,n_{an},n_{b1},\ldots,n_{bm}\rangle &=& \frac{1}{\sqrt{\prod_{j=1}^n n_{aj}!\prod_{k=1}^m n_{bk}!}}\prod_{j=1}^n (\hat{a}_{j}^{\dagger})^{n_{aj}}\prod_{k=1}^m(\hat{b}_{k}^{\dagger})^{n_{bj}}|0\rangle, \nonumber\\
\label{state1}   
\end{eqnarray}

\noindent where $|0\rangle = |0_{a1},\ldots,0_{an},0_{b1},\ldots,0_{bm}\rangle$ is the vacuum vector state in the Fock space. We can use the states (\ref{state1}) to write the matrix representation of the Hamiltonian (\ref{H1}). The dimension of the space increase very fast when we increase $N$,

$$	d = \frac{(n + m -1 + N)!}{(n + m -1)!N!}, $$

\noindent where $n + m$ is the total number of BECs in the system and $N$ is a constant $c$-number, $N = \sum_{j=1}^n n_{aj} + \sum_{k=1}^m n_{bk}$. In the case where we have only two BECs \cite{GSantosaa} (one $a_1$ and one $b_1$) the dimension is $d = N + 1$. In the Figs. (\ref{GF1}) and (\ref{GF2}) we show some graphs for different values of $n$ and $m$. The balls with their respective labels are representing the condensates and the tubes are representing the tunnelling of the atoms between the respective condensates. 

Now we use the co-multiplication property of the Lax operators to write,

\begin{equation}
  \hat{L}(u) = \hat{L}_{1}^{\Sigma_a^n}(u + \sum_{j=1}^{n}\omega_{aj})\hat{L}_{2}^{\Sigma_b^m}(u - \sum_{k=1}^{m}\omega_{bk}).
\label{LH2}
\end{equation}
\noindent Following the monodromy matrix (\ref{monod}) we can write the operators,

\begin{eqnarray}
\pi(\hat{A}(u)) &=& \left(u\hat{I} + \hat{I}\sum_{j=1}^n \omega_{aj} + \eta\sum_{j=1}^n \hat{N}_{aj}\right)\left(u\hat{I}  - \hat{I}\sum_{k=1}^m \omega_{bk} + \eta\sum_{k=1}^m \hat{N}_{bk}\right) \nonumber \\
          &+& \sum_{j=1}^n\sum_{k=1}^m t_{aj}s_{bk} \hat{b}_k^{\dagger}\hat{a}_j, \label{piAn}\\
\pi(\hat{B}(u)) &=& \left(u\hat{I} + \hat{I}\sum_{j=1}^n \omega_{aj} + \eta\sum_{j=1}^n \hat{N}_{aj}\right)\left(\sum_{k=1}^m t_{bk}\hat{b}_k\right) + \eta^{-1}\zeta_b\sum_{j=1}^n t_{aj}\hat{a}_j, \label{piBn}\\
\pi(\hat{C}(u)) &=& \left(\sum_{j=1}^n s_{aj}\hat{a}_j^{\dagger}\right)\left(u\hat{I}  - \hat{I}\sum_{k=1}^m \omega_{bk} + \eta\sum_{k=1}^m \hat{N}_{bk}\right) \nonumber \\
          &+& \eta^{-1}\zeta_a\sum_{k=1}^m s_{bk}\hat{b}_k^{\dagger}, \label{piCn}\\
\pi(\hat{D}(u)) &=& \sum_{j=1}^n\sum_{k=1}^m s_{aj}t_{bk}\hat{a}_j^{\dagger}\hat{b}_k + \eta^{-2}\zeta_a\zeta_b \hat{I}. \label{piDn}
\end{eqnarray}

Taking the trace of the operator (\ref{LH2}) we get the transfer matrix

\begin{eqnarray}
\hat{t}(u) & = & u^2 \hat{I} + u\hat{I}\left(\sum_{j=1}^n \omega_{aj}-\sum_{k=1}^m \omega_{bk}\right) + u\eta\hat{N} + \left(\eta^{-2}\zeta_a\zeta_b - \sum_{j=1}^n \sum_{k=1}^m \omega_{aj}\omega_{bk}\right)\hat{I}  \nonumber\\
	 & - & \eta\left(\sum_{k=1}^m \sum_{j=1}^n \omega_{bk}\hat{N}_{aj} - \sum_{j=1}^n \sum_{k=1}^m \omega_{aj}\hat{N}_{bk}   \right)  + \eta^2\sum_{j=1}^n \sum_{k=1}^m \hat{N}_{aj}\hat{N}_{bk}  \nonumber \\
 	 & + & \sum_{j=1}^{n}\sum_{k=1}^{m}(s_{aj}t_{bk}\hat{a}_{j}^{\dagger}\hat{b}_{k} + s_{bk}t_{aj}\hat{b}_{k}^{\dagger}\hat{a}_{j}).
\label{tu2} 
\end{eqnarray}

\noindent From (\ref{C14b}) we identify the conserved quantities of the transfer matrix (\ref{tu2}),

\begin{eqnarray}
\hat{\mathcal{C}}_0 
	 & = &  \left(\eta^{-2}\zeta_a\zeta_b - \sum_{j=1}^n \sum_{k=1}^m \omega_{aj}\omega_{bk}\right)\hat{I}  \nonumber\\
	 & - & \eta\left(\sum_{k=1}^m \sum_{j=1}^n \omega_{bk}\hat{N}_{aj} - \sum_{j=1}^n \sum_{k=1}^m \omega_{aj}\hat{N}_{bk}   \right)  + \eta^2\sum_{j=1}^n \sum_{k=1}^m \hat{N}_{aj}\hat{N}_{bk}  \nonumber \\
 	 & + & \sum_{j=1}^{n}\sum_{k=1}^{m}(s_{aj}t_{bk}\hat{a}_{j}^{\dagger}\hat{b}_{k} + s_{bk}t_{aj}\hat{b}_{k}^{\dagger}\hat{a}_{j}),
\end{eqnarray}

\begin{eqnarray}
\hat{\mathcal{C}}_1 &=&  \hat{I}\left(\sum_{j=1}^n \omega_{aj}-\sum_{k=1}^m \omega_{bk}\right) + \eta\hat{N},
\end{eqnarray}

\begin{eqnarray}
\hat{\mathcal{C}}_2 &=& \hat{I}.
\end{eqnarray}
We can rewrite the Hamiltonian (\ref{H1}) using these conserved quantities

\begin{equation}
 \hat{H} = \xi_0\hat{\mathcal{C}}_0  + \xi_1\hat{\mathcal{C}}_1^2 + \xi_2\hat{\mathcal{C}}_2,
\label{H4} 
\end{equation}
with the following identification for the parameters
\begin{eqnarray}
\xi_2 &=& - \xi_0\left( \eta^{-2} \zeta_a\zeta_b - \omega_{ab}^{nm}\right) - \xi_1\Delta\omega_{ab}^2,\\
\Delta\omega_{ab} &=& \sum_{j=1}^n \omega_{aj} - \sum_{k=1}^m \omega_{bk}, \\
\omega_{ab}^{nm}  &=& \sum_{j=1}^n\sum_{k=1}^m \omega_{aj}\omega_{bk},
\end{eqnarray}

\begin{equation}
U_{ajbk} = \eta^2(\xi_0 + 2\xi_1), \qquad U_{ajak} =U_{bjbk} = \eta^2\xi_1,
\end{equation}

\begin{equation}
\epsilon _{aj} - \sum_{k=1}^m\mu_{ajbk} = 2\eta\xi_1\Delta\omega_{ab} - \eta\xi_0\sum_{k=1}^m \omega_{bk},
\end{equation}

\begin{equation}
\epsilon _{bk} + \sum_{j=1}^n\mu_{ajbk} = 2\eta\xi_1\Delta\omega_{ab} + \eta\xi_0\sum_{j=1}^n \omega_{aj},
\end{equation}

\begin{equation}
J_{ajbk} = -\xi_0 t_{aj} s_{bk} = -\xi_0 s_{aj} t_{bk}.
\end{equation}

The Hamiltonian (\ref{H1}) is related with the transfer matrix (\ref{tu2}) by the equation,

\begin{equation}
 \hat{H} = \xi_0\hat{t}(u) + \xi_1\hat{\mathcal{C}}_1^2 - \xi_0\hat{\mathcal{C}}_1u - (\xi_0 u^2 - \xi_2)\hat{\mathcal{C}}_2.
\label{H5} 
\end{equation}

We use as pseudo-vacuum the product state, 

\begin{equation}
|0\rangle = \left(\bigotimes_{j=1}^n|0\rangle_{a_j}\right)\otimes \left(\bigotimes_{k=1}^m|0\rangle_{b_k}\right),
\end{equation}
\noindent with $|0\rangle_{a_j}$ denoting the Fock vacuum state for the BECs $a_j$ and $|0\rangle_{b_k}$ denoting the Fock vacuum state for the BECs $b_k$, for $j=1,\ldots, n$ and $k=1\ldots, m$. For this pseudo-vacuum we can apply the algebraic Bethe ansatz method in order to find the Bethe ansatz equations (BAEs),

\begin{eqnarray}
\frac{v^2_{i} + v_{i}\Delta\omega_{ab} - \omega_{ab}^{nm}}{\eta^{-2}\zeta_a\zeta_b} & = & \prod_{j \ne i}^{N}\frac{v_{i}-v_{j}-\eta}{v_{i}-v_{j}+\eta}, \;\;\;\;\;  i,j = 1,\ldots , N. \nonumber\\
\label{BAE2}
\end{eqnarray}

The eigenvectors \cite{Bethe-states}  $\{ |v_1,v_2,\ldots,v_N\rangle \}$ of the Hamiltonian (\ref{H1}) or (\ref{H4}) and of the transfer matrix (\ref{tu2}) are 

\begin{equation}
|\vec{v}\rangle \equiv  |v_1,v_2,\ldots,v_N\rangle = \prod_{i=1}^N \pi(\hat{C}(v_i))|0 \rangle,
\end{equation}
\noindent and the eigenvalues of the Hamiltonian (\ref{H1}) or (\ref{H4}) are,

\begin{eqnarray}
E(\{ v_i \}) & = &   \xi_0\left(u^2 + u\Delta\omega_{ab} - \omega_{ab}^{nm}\right) \prod_{i=1}^{N}\frac{v_{i} - u +\eta}{v_{i} - u} \nonumber \\
  & + &  \xi_0\eta^{-2}\zeta_a\zeta_b\prod_{i=1}^{N}\frac{v_{i} - u -\eta}{v_{i} - u} + \xi_1\mathcal{C}_1^2 - \xi_0\mathcal{C}_1u - \xi_0 u^2 + \xi_2,
\end{eqnarray}
\noindent where the $\{v_i\}$ are solutions of the BAEs (\ref{BAE2}) and $N$ is the total number of atoms. We can choose arbitrarily the spectral parameter $u$.

Choosing $\frac{J_{ajbk}}{\xi_0} = - \eta$, for example, we can write the BAEs (\ref{BAE2}) in the limit $\eta \rightarrow 0$ as  just one equation
\begin{eqnarray}
\sum_{i=1}^N \left(v_{i} + \frac{1}{2}\Delta\omega_{ab} \right)^2  & = & R_N^2,  \nonumber\\
\label{BAE3}
\end{eqnarray}  
\noindent with

\begin{equation}
R_N = \sqrt{\left(\frac{1}{4}\Delta\omega^2_{ab} + \omega_{ab}^{nm} + n\times m \right) N}.
\end{equation}
\noindent If the Bethe roots $\{v_i\}$ are real numbers, the BAE (\ref{BAE3}) is the equation of a $N$-dimensional sphere of radii $R_N$ and center in 

\begin{equation}
v_{i} = - \frac{1}{2}\Delta\omega_{ab}, \qquad \forall \; i = 1, \ldots, N.
\end{equation}

In this limit and with $u = 0$ we can write the eigenvalues as

\begin{eqnarray}
E(\{ v_i \}) & = &   \xi_1\left(\mathcal{C}_1^2  - \Delta\omega_{ab}^{2}\right)  -  \eta\xi_0\left(n\times m + \omega_{ab}^{nm}\right)\sum_{i=1}^{N} \frac{1}{v_{i}}.  
\end{eqnarray}

\section{Summary}
I have solved a family of fragmented Bose-Hubbard models using the multi-states  boson Lax operators introduced by the author in \cite{multistates}. These models can be considered as graphs, with the BECs in the vertices and the edge representing the tunnelling between the respective BECs. The graphs can appear in one-dimension (1D), two-dimension (2D) and in three-dimension (3D). When we increase the number of BECs we get a ring of BECs $a$ and a chain of BECs $b$ in the center of the ring. We can consider the BECs identical or different. I have showed that in the limit $\eta \rightarrow 0$, if the Bethe roots are all real numbers, they are on a $N$-dimensional sphere.

\section*{Acknowledgments}
The author acknowledge Capes/FAPERJ (Coordena\c{c}\~ao de Aperfei\c{c}oamento de Pessoal de N\'{\i}vel Superior/Funda\c{c}\~ao de Amparo \`a Pesquisa do Estado do Rio de Janeiro) for financial support.

\end{document}